# BRST Invariant Theory Of A Generalized 1+1 Dimensional Nonlinear Sigma Model With Topological Term

Yong-Chang Huang[1,3]   Kai-Hua Yang[1]   Xi-Guo Lee[2]

1. Institute of Theoretical Physics, Beijing University of Technology, Beijing, 100022, P. R. China
2. Institute of Modern Physics, Chinese Academy of Sciences, Lanzhou, 730000, P.R. China
3. CCAST ( World Lab. ), P. O. Box 8730, Beijing, 100080, P. R. China

## Abstract

We give a generalized Lagrangian density of 1+1 Dimensional O(3) nonlinear $\sigma$ model with subsidiary constraints, different Lagrange multiplier fields and topological term, find a lost intrinsic constraint condition, convert the subsidiary constraints into inner constraints in the nonlinear $\sigma$ model, give the example of not introducing the lost constraint $\dot{N} = 0$, by comparing the example with the case of introducing the lost constraint, we obtain that when not introducing the lost constraint, one has to obtain a lot of various non-intrinsic constraints. We further deduce the gauge generator, give general BRST transformation of the model under the general conditions. It is discovered that there exists a gauge parameter $\beta$ originating from the freedom degree of BRST transformation in a general O(3) nonlinear sigma model, and we gain the general commutation relations of ghost field.

## 1. Introduction

1+1 dimensional nonlinear $\sigma$ - model was introduced, at the beginning, as effective theories describing the interaction of Goldstone particles [1], these theories have the many similar points with 3+1 dimensional Yang- Mills theories. And the two are both the scale invariance and renormalizable [2]. Another important similar property is that they are both constraint systems: in Yang- Mills theories, constraints are acquired due to the gauge freedom degrees; in $\sigma$ - model, usually adopting the different method, e.g., taking the Lagrange multiplier fields. And noncommutative Yang-Mills systems [3] are still constraint systems. 1+1 dimensional nonlinear $\sigma$ - model has instantons, asymptotic freedom and nonperturbative spectra, which are analogous to gauge field theories in 3+1 dimensional spacetime [4]. Panigrahi, Roy, Scherer, Wilczek, Wu and Zee, extensively studied 2+1 dimensional different nonlinear $\sigma$ - models with Hopf terms or Chern-Simons [5,6], and Dzyaloshinskii, Polyakov and Wiegmann used nonlinear $\sigma$ - models to research high temperature superconductivity and fractional Hall effect [7,8].

Gauge symmetry enhancement and radiatively induced mass in the large N nonlinear sigma model are given [9], Ref.[10] researched granular superconductors: from the nonlinear sigma model to the Bose-Hubbard description, and Kamenev presented weak charge quantization as an instanton of the interacting sigma model

[11].

BRST invariant theories are the powerful tools studying the renormalizable standard model [12-14], and the theories have been generally applied to both general gauge theories [15] and string theories [16].

Dirac theory [17] is a well-known theory quantizing constraint physical systems, the Poisson brackets in second-class constraint systems are transformed into Dirac brackets to make the system solvable. But the Dirac brackets have ordering problem of field operators. When it is possible to convert a second-class constraint system into a first-class constraint system, one can use Poisson brackets to achieve the corresponding quantum commutators. Amorim, Barcelos-Neto, Boschi-Filho, Ghosh, Henneaux, Kim, Nativiade, Park, Rothe and Wilch well researched the current interests about nonlinear σ models [18,19].

This letter develops a method converting the second-class constraints into first class ones by introducing auxiliary constraints and fields, and this formalism may be used to all nonlinear σ models, and this approach may also be applied to $CP_1$ model [20,21] and so on different nonlinear σ models.

## 2. Hamiltonian description of a generalized 1+1 dimensional O(3) nonlinear σ model with topological term

A generalized Lagrangian density of 1+1 dimensional O(3) nonlinear σ model with subsidiary constraints, different Lagrange multiplier fields and topological term is

$$\mathcal{L} = \frac{1}{2q^2}(\partial_\mu N^i)(\partial^\mu N^i) + \frac{k}{4\pi}\varepsilon^{\mu\nu}\varepsilon^{ijk}N^i\partial_\mu N^j\partial_\nu N^k - \lambda_0(N^iN^i - 1) - (\partial_\mu\lambda_1)(\partial^\mu N), \quad (1)$$

where q is a coupling constant; μ = 0, 1; i = 1, 2, 3; $\varepsilon^{01} = -\varepsilon^{10} = 1$, $\varepsilon^{123} = -\varepsilon^{213} = 1$ and $(N)^2 = N^iN^i$; $\lambda_0$, $\dot\lambda_1$ and $\lambda'_1$ are different characteristic multiplier fields relative to subsidiary constraints $(N)^2 - 1 = 0$, $\dot N = 0$ and $N' = 0$, respectively. Because of $N = \pm 1$, one may equivalently take N = 1 as constraint in Eq.(1), and in the standard polar coordinate, since the action contains a full integration over the full solid angle, the sign of the variable R can be changed at will by a redefinition $\theta \to \pi - \theta, \phi \to \pi + \phi$, therefore, this is totally consistent. The Lagrangian is still nonlinear (see M. E. Peskin and D. V. Schroeder, An introduction to quantum field theory, Addison-Wesley publishing Company, 1995, P. 455), so does the nonlinear topological term. $(\partial_\mu\lambda_1)(\partial^\mu N)$ in Eq.(1) still satisfies the inverse symmetries of time and space. It is well known from constraint theory that the Lagrange multiplier field $\lambda_1$ is only time function, but $\lambda_1$ isn't only space coordinate function, accordingly, $(\partial_\mu\lambda_1)(\partial^\mu N)$ is simplified as $\dot\lambda_1 \dot N$. In terns of constraint theory, using a multiplier field to multiply a constraint condition and adding the product into the Lagrangian mean that both extending variables and temporarily loosing constraint condition, especially, $\dot N = N^i\partial_0 N^i = 0$ means that different components $N^i$ and $\partial_0 N^i$, respectively, of $N$ and $\dot N$ are variable with time t evolution, but they need to satisfy a constraint equation $N^i\partial_0 N^i = 0$. Because the constraint equation is a natural result from N = 1, this kind of systems is very many in particle physics, field theory, condensed physics etc, thus this letter's studies are useful.

Because the system has O(3) symmetry, we may use polar coordinates to represent the fields

$$N^1 = R\sin\theta\cos\phi, \quad N^2 = R\sin\theta\sin\phi, \quad N^3 = R\cos\theta. \quad (2)$$

Through a very long calculation, Eq.(1) can be rewritten as

$$\mathcal{L} = \frac{1}{2q^2}\{\partial_\mu R\partial^\mu R + R^2[\partial_\mu\theta\partial^\mu\theta + \sin^2\theta\partial_\mu\phi\partial^\mu\phi]\} - \lambda_0(R-1) - \dot\lambda_1\dot R$$

$$+kR^3\sin\theta\,(\partial_0\theta\,\partial_1\phi-\partial_0\phi\,\partial_1\theta)/2\pi \quad . \tag{3}$$

In the new field variables, the canonical momenta are

$$\pi_R=\frac{\partial\mathcal{L}}{\partial(\partial_0 R)}=\frac{\dot{R}}{q^2}-\dot{\lambda}_1, \quad \pi_\theta=\frac{R^2\dot\theta}{q^2}+\frac{k}{2\pi}R^3\sin\theta\,\partial_1\phi \tag{4}$$

$$\pi_\phi=\frac{R^2\sin^2\theta\,\dot\phi}{q^2}-\frac{k}{2\pi}R^3\sin\theta\,\partial_1\theta \quad , \quad \pi_{\lambda_0}=0, \quad \pi_{\lambda_1}=-\dot{R}, \tag{5}$$

the second equation in (5) is a primary constraint in the system of polar coordinates. Therefore, in Eqs.(1&3), we have converted the subsidiary constraints into inner constraints in the nonlinear σ model.

Accordingly, in phase space, Eq.( 3) can be expressed as

$$\mathcal{L}_p=\frac{\pi_{\lambda_1}^2}{2q^2}-\frac{1}{2q^2}(\partial_1 R)^2+\frac{q^2}{2R^2}\left(\pi_\theta-\frac{k}{2\pi}R^3\sin\theta\,\partial_1\phi\right)^2-\frac{R^2}{2q^2}(\partial_1\theta)^2+$$
$$\frac{q^2}{2R^2\sin^2\theta}\left(\pi_\varphi+\frac{k}{2\pi}R^3\sin\theta\,\partial_1\theta\right)^2-\frac{R^2\sin^2\theta}{2q^2}(\partial_1\varphi)^2-\lambda_0(R-1)$$
$$+\frac{k}{2\pi}R^3\sin\theta\,(\partial_0\theta\,\partial_1\phi-\partial_0\phi\,\partial_1\theta)\;-(\frac{\pi_{\lambda_1}}{q^2}+\pi_R)\pi_{\lambda_1}, \tag{6}$$

we, thus, can obtain the corresponding Hamiltonian density in phase space as follows

$$\mathcal{H}_p=\pi_R\dot{R}+\pi_\theta\dot\theta+\pi_\pi\dot\varphi+\pi_{\lambda_0}\dot\lambda_0+\pi_{\lambda_1}\dot\lambda_1-\mathcal{L}_p$$
$$=\frac{1}{2}\left\{-\frac{\pi_{\lambda_1}^2}{q^2}+\frac{q^2}{R^2}\left[\pi_\theta^2+\frac{1}{\sin^2\theta}\pi_\phi^2\right]\right\}+\frac{1}{2q^2}\left\{(\partial_1 R)^2+R^2\left(1+\frac{k^2 g^4}{4\pi^2}R^2\right)\left[(\partial_1\theta)^2+\right.\right.$$
$$\left.\left.\sin^2\theta(\partial_1\phi)^2\right]\right\}+\frac{kq^2}{2\pi}R\left(\frac{1}{\sin\theta}\pi_\phi\partial_1\theta-\sin\theta\pi_\theta\partial_1\phi\right)+\lambda_0(t)(R-1)-\pi_R\pi_{\lambda_1}. \tag{7}$$

Therefore, we obtain the total Hamiltonian density $\mathcal{H}_T=\mathcal{H}_p+\mu_0\pi_{\lambda_0}$, $\mu_0$ is Lagrange multiplier, because we have converted the subsidiary constraints into inner constraint and make the two subsidiary constraints change as an inner primary constraint in the nonlinear σ model, which make the model easier to deal with, see the following section.

## 3. Gauge fixing generalization, gauge generator, gauge invariance and deduction of general BRST transformation

Using the Lagrangian and Hamiltonian densities of the above system, we deduce general BRST transformation by introducing a general fixed item and gauge generator.

Utilizing Dirac-Bergman constrained theory [22] and the primary constraint

$$\Phi_0=\pi_{\lambda_0}=0 \quad , \tag{8}$$

we deduce the secondary constraints as follows

$$\Phi_1=1-R \quad , \quad \Phi_2=\pi_{\lambda_1} \quad , \tag{9}$$

there are not the other constraints after calculation, $\Phi_0$, $\Phi_1$ and $\Phi_2$ are all the constraints of the first class. *If we do not introduce the subsidiary constraint $\dot{N}=0$, we have to obtain and have difficultly to deal with a lot of various constraints* in the usual constraint theory [22], for example, when *not introducing the constraint $\dot{N}=0$, one has to obtain a lot of various constraints* $\phi_0=\pi_{\lambda_0}$, $\phi_1=1-R$, $\phi_2=-q^2\pi_R$, $\phi_3 = q^2 \partial H / \partial R$, $\phi_4 = \{q^2 \partial H / \partial R, H\}$, ......, because $\phi_4$ has many terms, which may further yield a lot of various constraints, these constraints are not intrinsic, and there are the first class and the second class in these constraints, under the cases, it is very difficult that one wants to find out general BRST transformation and to quantize the theory.

Because we have cancelled the superabundant non-intrinsic constraints and have converted the all constraints into the constraints of the first class, in terms of the definition of gauge generator[22], we can have the gauge generator that can generally deduce BRST transformation as follows

$$G = \int [\ddot{\varepsilon}\pi_{\lambda_0} + \dot{\varepsilon}(R-1) + \varepsilon \pi_{\lambda_1}] dx \quad , \tag{10}$$

Using the gauge generator G we can obtain the following gauge transformation

$$\delta R = \{R, G\} = 0, \quad \delta\theta = \{\theta, G\} = 0, \quad \delta\phi = \{\varphi, G\} = 0, \quad \delta\lambda_0 = \{\lambda_0, G\} = \ddot{\varepsilon},$$

$$\delta\pi_R = \{\pi_R, G\} = -\dot{\varepsilon}, \quad \delta\pi_\theta = \{\pi_\theta, G\} = 0, \delta\pi_{\lambda_0} = \{\pi_{\lambda_0}, G\} = 0, \delta\pi_\phi = \{\pi_\phi, G\} = 0$$

$$\delta\lambda_1 = \{\lambda_1, G\} = \varepsilon, \quad \delta\pi_{\lambda_1} = \{\pi_{\lambda_1}, G\} = 0 \quad . \tag{11}$$

Define $\dot{\varepsilon}(t) = \omega c(t)$, where $\omega$ is a Grassmann number not depending time t, $c(t)$ is a Grassmann number, then $\dot{\varepsilon}(t)$ is a commutative number. Using the definition, we set up the connection between gauge transformation and BRST transformation, thus, we obtain

$$\delta\pi_R = -\omega c, \quad \delta\lambda_0 = \omega\dot{c}, \quad \delta\lambda_1 = \omega \int_{t_0}^{t} c\, dt \quad , \tag{12}$$

where, not losing generality, we have taken $\varepsilon(t_0) = 0$. For the generalized model, we take a generalized gauge fixed term $\mathcal{L}_{GF}$ of Lagrangian density that is different from all known gauge fixed terms as follows

$$\mathcal{L}_{GF} = b[\pi_R - (\beta+1)\dot{\lambda}_0] - \frac{1}{2}b^2 \quad , \tag{13}$$

where $\beta$ is an arbitrary constant number not equating to $-1$, b is an assistant commutative field. In order to make general research, we can further take a generalized Faddev-Poppov Lagrangian density $\mathcal{L}_{FP}$ different from all known Faddev-Poppov Lagrangian densities as follows

$$\mathcal{L}_{FP} = (\beta+1)\dot{\bar{c}}\dot{c} - \bar{c}c \quad , \tag{14}$$

Therefore, we obtain a new generalized effective Lagrangian density $\mathcal{L}_{eff} = \mathcal{L}_p + \mathcal{L}_{GF+FP}$, where $c$ is F-P ghost field, $\bar{c}$ is anti-ghost field. When $\beta = 0$, Eqs.(13&14) return to usual cases, and $\delta\mathcal{L}_{GF+FP} = 0$ satisfies invariance of BRST transformation, thus the generalizations are consistent.

Under the condition of keeping $\mathcal{L}_{eff}$ invariant, we can find out the transformations of fields of making ghost field and gauge fixed terms invariant, that is

$$\delta\mathcal{L}_{GF+FP} = \delta\left[b[\pi_R - (\beta+1)\dot{\lambda}_0] - \frac{1}{2}b^2 + (\beta+1)\dot{\bar{c}}\dot{c} - \bar{c}c\right]$$

$$= \delta b[\pi_R - (\beta+1)\dot{\lambda}_0 - b] - (b\omega + \delta\bar{c})[(\beta+1)\ddot{c} + c] + [(\beta+1)\ddot{\bar{c}} + \bar{c}]\delta c + (\beta+1)\frac{d}{dt}(\delta\bar{c}\dot{c} - \dot{\bar{c}}\delta c), \tag{15}$$

Due to $\delta\mathcal{L}_{GF+FP} = 0$ ( i.e., satisfying invariance of BRST transformation ), and the last term is the whole derivative term about time t, this term may be neglected when variational is done in integration. Accordingly we deduce when $\pi_R - (\beta+1)\dot{\lambda}_0 - b \neq 0$ and $(\beta+1)\ddot{c} + c \neq 0$, there are the following three transformations

$$\delta b = 0, \quad \delta\bar{c} = -b\omega, \quad \delta c = 0 \quad . \tag{16}$$

We finally deduce the general BRST transformation as follows

$$\delta R = \delta\theta = \delta\phi = \delta\pi_\theta = \delta\pi_{\lambda_0} = \delta\pi_{\lambda_1} = \delta\pi_\varphi = \delta c = \delta b = 0, \qquad (17)$$

$$\delta\bar{c} = -b\omega, \quad \delta\lambda_0 = \omega\dot{c}, \quad \delta\pi_R = -\omega c, \quad \delta\lambda_1 = \omega\int_{t_0}^{t} c\, dt \quad . \qquad (18)$$

Using the last formula in (18), we gain $\delta\dot{\lambda}_1 = \omega c$ in Eqs.(1) and (3).

Because $\omega$ is a Grassmann constant, we can again obtain the general BRST transformation

$$\delta\bar{c} = -b, \quad \delta\lambda_0 = \dot{c}, \quad \delta\pi_R = -c, \quad \delta\lambda_1 = \int_{t_0}^{t} c\, dt \quad . \qquad (19)$$

Thus, Eq.(18) is equivalent to Eq.(19). When taking $\beta(\beta \neq -1)$ different values, we achieve different conservation quantities and corresponding general conclusion, and when $\beta = 0$ or/and $\lambda_1 = 0$, the above all results are simplified, e.g., when divorcing $-\omega$ from Eqs.(18) the transformation is simplified as the results corresponding to $C_2$ constraint in Eq.(27) in Ref.[24], thus, our researches are general, and we discover that there is a gauge parameter $\beta$ in general BRST transformation in a general O(3) nonlinear sigma model, and the gauge parameter $\beta$ affects both the Lagrangian density and the Hamiltonian densities, thus, the gauge parameter $\beta$ may influences Euler-Lagrange equations, conservation quantities and quantization etc of this system, because the gauge parameter $\beta$ originates from freedom degree of BRST transformation, the gauge parameter $\beta$ is analogous to the gauge parameter originating from freedom degree of U(1) gauge transformation in electromagnetic field theory, then, the gauge parameter $\beta$ has physical meanings.

When there is not the $\dot{\lambda}_1 \dot{N}$ term in Lagrangian (1), then there is not the term $\dot{\lambda}_1 \dot{R}$ in Eq. (3), which result in that the first term in (9) and the other secondary constraints are the second class constraints, therefore, adding the term $\dot{\lambda}_1 \dot{N}$ into Lagrangian (1) makes the second class constraints automatically change as the first class constraints due to *the whole consistent inner structure* of the constraints and BRST transformation of the nonlinear sigma model, namely, which consistently converts the second-class constraints into the first class ones by introducing auxiliary constraint and field, and $\dot{\lambda}_1 \dot{N}$ ( not like the nonlinear sigma models in all articles and books ) guarantees the invariance of time-reverse of the Lagrangian density; because all the monomials in the Lagrangian are of dimension smaller than four, the theory is renormalizable [25], thus, we find the whole consistent inner structure of the nonlinear sigma model, this formalism may been directly or extendedly applied to almost all nonlinear σ models and $CP_1$ model etc, because these models have the lost constraint $\dot{N}$ and the studies relative to $\dot{\lambda}_1, \dot{N}$ etc have not been done up to now.

**4. Generalizing Lagrangian density of gauge fixing and ghost field and corresponding Hamiltonian density**

Using Lagrangian density $\mathcal{L}_p$ in phase space and Eqs.(13) and (14), we get the generalized Lagrangian density that is invariant under the general BRST transformation as follows

$$\mathcal{L}_{BRST} = \mathcal{L}_p + \mathcal{L}_{GF+FP}$$

$$= \frac{\pi_{\lambda_1}^2}{2q^2} - \frac{1}{2q^2}(\partial_1 R)^2 + \frac{q^2}{2R^2}\left(\pi_\theta - \frac{k}{2\pi}R^3 \sin\theta\, \partial_1\phi\right)^2 - \frac{R^2}{2q^2}(\partial_1\theta)^2 +$$

$$\frac{q^2}{2R^2\sin^2\theta}\left(\pi_\varphi + \frac{k}{2\pi}R^3\sin\theta\,\partial_1\theta\right)^2 - \frac{R^2\sin^2\theta}{2q^2}(\partial_1\varphi)^2 - \lambda(R-1) + \frac{k}{2\pi}R^3\sin\theta\cdot$$

$$(\partial_0\theta\,\partial_1\phi - \partial_0\phi\,\partial_1\theta) - (\frac{\pi_{\lambda_1}}{q^2} + \pi_R)\pi_{\lambda_1} + \left[b[\pi_R - (\beta+1)\dot\lambda_0] - \frac{1}{2}b^2 + (\beta+1)\dot{\bar c}\dot c - \bar c c\right], \quad (20)$$

where $\delta\mathcal{L}_{BRST} = \mathcal{L}_{GF+FP}$ satisfies the nilpotency $\delta^2\mathcal{L}_{BRST} = 0$, then the relative canonical momenta are

$$\pi_{\lambda_0} = -(\beta+1)b, \quad \pi_c = -(\beta+1)\dot{\bar c}, \quad \pi_{\bar c} = (\beta+1)\dot c \quad . \quad (21)$$

Therefore, the general Hamiltonian density containing ghost field and general gauge condition is

$$\mathcal{H}_{BRST} = \pi_R\dot R + \pi_\theta\dot\theta + \pi_\varphi\dot\varphi + \pi_{\lambda_0}\dot\lambda_0 + \pi_{\lambda_1}\dot\lambda_1 + \pi_c\dot c + \dot{\bar c}\pi_{\bar c} - \mathcal{L}_{BRST}$$

$$= \frac{1}{2}\left\{-\frac{\pi_{\lambda_1}^2}{q^2} + \frac{q^2}{R^2}\left[\pi_\theta^2 + \frac{1}{\sin^2\theta}\pi_\varphi^2\right]\right\} + \frac{1}{2q^2}\left\{(\partial_1 R)^2 + R^2\left(1 + \frac{k^2 g^4}{4\pi^2}R^2\right)[(\partial_1\theta)^2 + \right.$$

$$\left.\sin^2\theta(\partial_1\phi)^2]\right\} + \frac{kq^2}{2\pi}R\left(\frac{1}{\sin\theta}\pi_\varphi\partial_1\theta - \sin\theta\pi_\theta\partial_1\phi\right) + \lambda(t)(R-1) - \pi_R\pi_{\lambda_1}$$

$$+ \frac{1}{(\beta+1)}\pi_c\pi_{\bar c} + \frac{1}{2(\beta+1)^2}\pi_{\lambda_0}^2 + \frac{1}{(\beta+1)}\pi_{\lambda_0}\pi_R + \bar c c \quad , \quad (22)$$

where the whole differential term about time t has be neglected.

In the following, we find out their commutative relations. Using Eq.(22) we obtain the canonical equations of $c$ and $\bar c$ as follows

$$\{c, \mathcal{H}_{BRST}\} = \partial_0 c = \frac{\partial\mathcal{H}_{BRST}}{\partial\pi_c} = \frac{\pi_{\bar c}}{(\beta+1)}, \quad (23)$$

$$\{\bar c, \mathcal{H}_{BRST}\} = \partial_0 \bar c = \frac{\partial\mathcal{H}_{BRST}}{\partial\pi_{\bar c}} = -\frac{\pi_c}{(\beta+1)} \quad . \quad (24)$$

then Eqs.(21), (23 & 24) are consistent about $c$ and $\bar c$, thus, this letter's generalizations are consistent.

Because $c$ and $\bar c$ are both independent canonical variables, one may obtain that $\partial_0 c$ and $\bar c$ or $\partial_0\bar c$ and $c$ have anti-commutative relations, namely,

$$\{\pi_c, \pi_{\bar c}\} = \{c, \bar c\} = 0, \quad \partial_0\{\bar c, c\} = 0, \quad \{\partial_0\bar c, c\} = (-1)\{\partial_0 c, \bar c\} \quad , \quad (25)$$

$c$ satisfies also Heisenbeger moving equation of Fermi field

$$\{c, \mathcal{H}_{BRST}\} = i\partial_0 c = \frac{i\pi_{\bar c}}{(\beta+1)}, \quad (26)$$

which is also consistent with Eqs.(21) and (23), the other commutative relations can be similarly obtained.

On the other hand, due to $\{c(x), \pi_c(x')\} = i\delta(x-x')$, then we obtain

$$\{\bar c(x), \dot c(x')\} = \frac{i\delta(x-x')}{(\beta+1)} = -\{c(x), \dot{\bar c}(x')\} \quad , \quad (27)$$

Eq.(27) represents two new general anticommutative relations about ghost field, the minus sign in Eq.(27) is not trivial. When $\beta = 0$, Eq.(4.8) returns to the past anti-commutative relation. Because general physical processes should satisfy quantitative causal relation [26-28], e.g., Ref.[29] uses the no-loss-no-gain homeomorphic map transformation satisfying the quantitative causal relation to gain exact strain tensor formulas in Weitzenböck manifold, and due to action of the classical constraint conditions, when quantizing these constraints, there exist the corresponding effects, and because our researches not only are very general, but also are different from the past works, e.g., Refs.[30,31], we can finish not only the general BRST quantization of the system, but also give their relative conservation charge etc all more studies on different nonlinear sigma models, owing to space limit, these and a lot of applications and generalizations will be written in the other papers.

5. **Summary and conclusion**

This letter gives a generalized Lagrangian density of 1+1 Dimensional O(3) nonlinear σ model with subsidiary constraints, Lagrange multiplier fields and topological term, finds a lost intrinsic constraint condition, converts the subsidiary constraints into inner constraints in the nonlinear σ model, and makes the two subsidiary constraints change as the inner primary constraint in the nonlinear σ model, which make the model easier to deal with. The example of not introducing the lost constraint $\dot{N} = 0$ is given, by comparing the example with the case of introducing the lost constraint, we obtain that when not introducing the lost constraint, one has to obtain a lot of various non-intrinsic constraints. This letter gives both a generalized gauge fixed term $\mathcal{L}_{GF}$ and a generalized Faddev-Poppov Lagrangian density $\mathcal{L}_{FP}$, the both are different from all known ones. Using Dirac constrained theory and the extended condition, we deduce the gauge generator that can generally deduce BRST transformation, set up the connection between gauge transformation and BRST transformation, give the general BRST transformation of (1+1) Dimensional O(3) nonlinear model with topological term under the general conditions, and it is discovered that there exists a gauge parameter $\beta$ originating from freedom degree of BRST transformation in a general O(3) nonlinear sigma model, the gauge parameter $\beta$ influences Euler-Lagrange equations, conservation quantities and quantization etc of this system, and the gauge parameter $\beta$ is analogous to the gauge parameter originating from freedom degree of U(1) gauge transformation in electromagnetic field theory, then, the gauge parameter $\beta$ has important physical meaning. We gain the general commutation relations of ghost field, consistently convert the second-class constraints into first class ones by introducing auxiliary constraints and fields, and the theory is renormalizable, we find out the whole consistent inner structure of the constraints and BRST transformation of nonlinear sigma model, this formalism may be generally applied to some different kinds of nonlinear σ models.

**Acknowledgement:** the author is grateful for Prof. Z. P. Li and Y. S. Wu for discussion.
The work is supported by Beijing novel program and beijing excellent fund ( 20042D0501516 )

# References

[1] S. Coleman, J. Weiss and B. Zumino, Phys. Rev., 177(1961)2239; C. Callan, S. Coleman , J. Weiss and B. Zumino, Phys. Rev., 177(1969)2247.
[2] A. I. Vainshtein, V. I. Zakharov, V. A. Novikov and M. A. Shifman, Sov. J. Pant , Nucl., 17(1986)204; A. B. Zamolodchikov, Ann. Phys. (NY)., 120(1979)253.
[3]M. Li and Y. S. Wu, Phys. Rev. Lett., 84 (2000) 2084.

[4] A. M. Polyakov, Phys. Lett., B 59(1975)79; V. A. Novikv et al, Phys. Rep. 116(1984) 103.

[5] F. Wilczek, A. Zee, Phys. Rev. Lett. 51(1983)2250; M. J. Bowick, D. Karabali, L. C. R. Wijewardhana, Nucl. Phys. B271(1986)417; and references therein.

[6] Y. S. Wu, A. Zee: Phys. Lett. 147B (1984) 325; P. Panigrahi, S. Roy, W. Scherer, Phys. Rev. D38(1988)3199; Phys. Rev. Lett. 61(1988)2827.

[7] A. M. Polyakov, Mod. Phys. Lett. A3 (1988) 455; I. Dzyaloshinskii, A.M. Polyakov, P. Wiegmann, Phys. Lett. A127(1988) 112.

[8] P. B. Wiegmann, Phys. Rev. Lett. 60(1988)821.

[9] T. Itoh, P. Oh and C. Ryou, Physical Review, D 64 (2001) 045005/1-5.

[10] I.V. Yurkevich and I.V. Lerner, Physical Review, B 64(2001)054515/1-5.

[11] A. Kamenev, Phys. Rev. Lett. 85,19 (2000)4160.

[12] C. Becchi, A. Rouet and R. Stora, Comm. Math. Phys. 42 (1975) 127; Ann. Phys. 98 (1976) 287.

[13] G.t'Hooft, Nucl. Phys. B33 (1971) 173; Nucl. Phys. B35 (1971) 167; G. t'Hooft and M. Veltman, Nucl. Phys. B44(1972) 189; B.W. Lee and J. Zinn-Justin, Phys. Rev. D5 (1972) 3121; D7 (1973) 1049; B.W. Lee, Phys. Rev. D9 (1974) 938.

[14] T. Kugo and I. Ojima, Prog. Theor. Phys. Suppl. 66 (1979) 1.

[15] E.S. Fradkin and G.A. Vilkovisky, Phys. Lett. B55 (1975) 224; I.A. Batalin and G.A. Vilkovisky, Phys. Lett. B69 (1977) 309.

[16] M. Kato and K. Ogawa, Nucl. Phys. B212 (1983) 443; S. Hwang, Phys. Rev. D28 (1983) 2614; W. Siegel, Phys. Lett. B149 (1984) 157, 162.

[17] P. A. M. Dirac, Lectures in Quantum Mechanics (Yeshiva University, New York, 1964).

[18] R. Banerjee, Phys. Rev. D48, R5467 (1993); W.T. Kim and Y.-J. Park, Phys. Lett. B336 (1994) 376; Marc Henneaux and André Wilch, Phys. Rev. D58 (1998) 025017.

[19] Y.-W. Kim, Y.-J. Park, and K.D. Rothe, J. Phys. G24, 953 (1998); Y.-W. Kim and K. D. Rothe, Nucl. Phys. B510, 511 (1998); S. Ghosh, Phys. Rev. D49, 2990 (1994); R. Amorim and J. Barcelos-Neto, Phys. Rev. D53, 7129 (1996); C.P. Nativiade and H. Boschi-Filho, Phys. Rev. D62, 025016 (2000).

[20] N. Banerjee, S. Ghosh, and R. Banerjee, Phys. Rev. D49, 1996 (1994).

[21] R. Banerjee, Phys. Rev. D49, 2133 (1994).

[22] Marc Henneaux and Claudio Teitelboim, Quantization of Gauge Systems, Princeton University Press, Princeton, New Jersey, 1992.

[23] D. M. Gitman and I.V. Tyutin, Quantization of Fields with Constraints, Berlin, Springer-Verlag, 1990.

[24] Purushotham Voruganti, Phys. Rev., D39(1989) 1179.

[25] Claude Itzykson, Jean-Bernard Zuber, Quantum Field Theory, Mcgraw-Hill International Book Company, New York, (1980).

[26] Y. C. Huang, F. C. Ma and N. Zhang, Mod. Phys. Lett., B**18** (2004) 1367

[27] Y. C. Huang and G. Weng, Commun. Theor. Phys., **44** (2005) 757.

[28] Y. C. Huang, X. G. Lee and M. X. Shao, Mod. Phys. Lett., A21(2006) 1107.

[29] Y. C. Huang and B. L. Lin, Phys. Lett., A299 (2002) 6449.

[30] D. Friedan, Phys. Rev. Lett., 45 (1980) 1057.

[31] D. H. Friedan, Ann. Phys. 163 (1985) 318.